\documentclass{elsart}

\usepackage{graphics}

\begin{document}

\begin{frontmatter}

\title{Measurement of the trailing edge of cosmic-ray track signals 
from a round-tube drift chamber}
\author[TUAT]{Miho~Abe},
\author[TUAT]{Tsuneo~Emura}, and
\author[KEK]{Shigeru~Odaka\thanksref{Correspond}}
\address[TUAT]{Faculty of Engineering, Tokyo University of Agriculture 
and Technology, Koganei, Tokyo 184-8588, Japan}
\address[KEK]{High Energy Accelerator Research Organization (KEK), 
Tsukuba, Ibaraki 305-0801, Japan}
\thanks[Correspond]{Corresponding Author. 
E-mail address: shigeru.odaka@kek.jp.}

\begin{abstract}
The trailing edge of tube drift-chamber signals for charged particles 
is expected to provide information concerning the particle 
passage time. 
This information may be useful for separating meaningful signals from 
overlapping garbage at high-rate experiments, 
such as the future LHC experiments.
We carried out a cosmic-ray test using a small tube chamber 
in order to investigate the feasibility of this idea.
We achieved a trailing-edge time resolution of 12 ns in rms 
by applying simple pulse shaping to eliminate a signal tail.
A comparison with a Monte Carlo simulation indicates the importance of
well-optimized signal shaping to achieve good resolution.
The resolution may be further improved with better shaping. 
\end{abstract}

\end{frontmatter}


\section{Introduction}

Tube drift chambers employing round tubes as the cathode electrode 
are frequently used as an important device for charged-particle 
tracking at large experiments, such as the Monitored Drift Tubes (MDT) 
in the muon detection system of the ATLAS experiment \cite{ATLAS} 
at the Large Hadron Collider (LHC) to be built at CERN. 
Because of their simple structure, 
tube drift chambers provide us with easiness in construction 
and calibration. 
This is a great advantage for constructing a large detector system.

Along with such an advantage, tube drift chambers have an undesirable 
nature in applications to very high-rate experiments, 
such as those at LHC. 
They require relatively long time intervals to collect 
all meaningful signals.
For example, this time interval ({\it i.e.}, the maximum drift time) 
is expected to be about 500 ns in the case of ATLAS-MDT. 
This is appreciably longer than the planned beam-crossing 
interval of LHC (25 ns). 
Hence, the event data will be contaminated by garbage signals produced 
by particles from neighboring beam-crossings.
In addition, since the environmental radiation tends to be severe 
at high-rate experiments, 
the data may suffer from continuous garbage produced by the radiation.
These contaminations may deteriorate the track-reconstruction 
capability of the detector.

The source of signals from drift chambers is ionization 
electrons produced by charged particles passing 
through the chamber gas volume.
The produced electrons drift towards an anode wire placed 
at the center to induce electrical signals 
via an avalanche process around the wire.
The first arriving electrons, produced near the closest approach 
to the anode wire, form the leading edge of the signals. 
The leading edge, therefore, provides track position information.

\begin{figure}
 \centering \leavevmode
 \scalebox{0.8}{\includegraphics{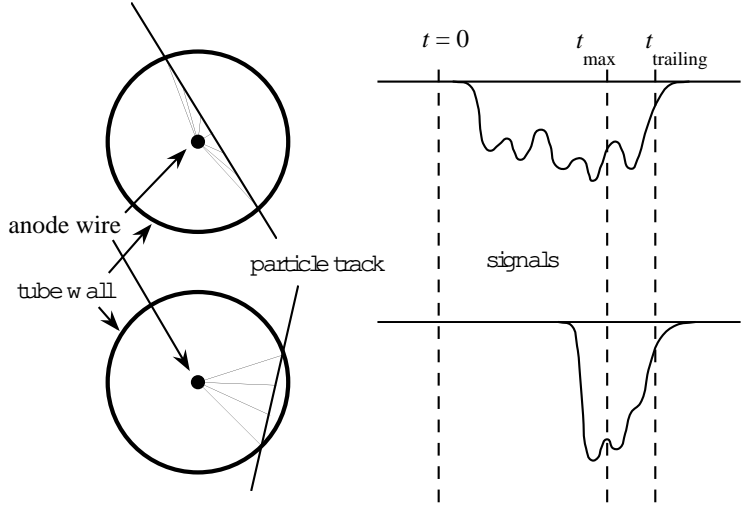}}
 \caption{Picture showing the idea. 
The trailing edge of tube chamber signals is expected to appear 
at an approximately identical time with respect to the particle 
passage, irrelevant of the leading-edge time and the incident angle.}
 \label{Fig_idea}
 \vspace{0.5cm}
\end{figure}

On the other hand, if the response to single electrons 
is sufficiently narrow, the trailing edge of the signals is 
determined by those electrons produced near the tube wall. 
Thus, in the case of round-tube chambers, 
trailing edges appear at an approximately identical time 
with respect to the charged-particle passage, 
irrelevant of the leading-edge time and the incident angle, 
as schematically shown in Fig.~\ref{Fig_idea}.

This argument leads to a prospect that,
if we can measure the trailing edge with a sufficient time resolution 
simultaneously with the leading edge, 
we may be able to distinguish the beam crossing relevant to each signal.  
In other words, we may be able to select only those signals relevant 
to an interesting beam crossing 
before applying reconstruction analyses \cite{idea}.

In order to investigate the feasibility of this idea,
we carried out a cosmic-ray test using a small tube drift chamber. 
The leading and trailing-edge times of the signals were simultaneously 
measured using a multi-hit TDC module 
employing the Time Memory Cell (TMC) LSI \cite{TMC-module}. 
We discuss the observed properties, by comparing the results 
with predictions from a Monte Carlo simulation.

\section{Setup for the cosmic-ray test}

The tube chamber used for the test is made of a thin aluminum tube 
having an inner diameter of 15.4 mm, a wall thickness of 0.1 mm, 
and a length of 20 cm.
A 3 mm-diameter window made on the tube wall near the center 
allows us to feed X rays into the tube.
The window is covered with a thin aluminum foil 
and sealed with Kapton adhesive tape.

A gold-plated tungsten wire of 30 $\mu$m in diameter is strung 
along the center axis of the tube.
A positive high voltage was applied to the wire 
with the tube wall being grounded.
A gas mixture of argon (50\%) and ethane (50\%) at the atmospheric 
pressure was made to flow inside the tube.

The signals from the tube chamber were amplified and discriminated 
using circuits made for the central drift chamber (CDC) of 
the VENUS experiment at KEK-TRISTAN, 
where a preamplifier board was attached to one end of the tube. 
The amplified signals were transferred to a discriminator board 
through a 30 m-long shielded twisted-pair cable.
We added a pulse shaping circuit to the discriminator board,
because no intentional shaping was applied in the original circuit. 

The timing of the discriminated signals was measured 
using a 32-ch TMC-VME module \cite{TMC-module} installed 
in a VME crate. 
This TDC module employs the TMC-TEG3 LSI \cite{TMC-TEG3}, 
allowing us to measure both the leading and trailing edges 
simultaneously with a time resolution of 0.37 ns.
The module also has an essentially unlimited multi-hit capability.

The TDC module was operated in a common-stop mode.
Stop signals synchronized with the passage of cosmic rays 
were formed by a coincidence between signals 
from two scintillation counters, 10 cm by 25 cm each, 
vertically sandwitching the tube chamber with a separation of 15 cm.
The coverage of this counter telescope was sufficiently larger than 
the tube chamber, providing a uniform track distribution 
in sampled data.
As a trade-off, tube-chamber signals were observed in only about 20\% 
of the data.

The data taking was controlled by a board computer installed 
in the VME crate.
The computer collected the digitized data and stored them 
in a local disk. 
After completing the data taking, 
the stored data were transferred to a workstation through a network 
and analyzed there.

\section{Pulse shaping}

Before starting the test, we naively thought that 
appropriate pulse shaping would be necessary 
for trailing-edge measurements with a good time resolution, 
because of the presence of a long $1/t$ tail in drift chamber signals.
The existence of a long tail, which may also be produced 
by readout circuits, would enhance the time-walk effect 
due to a large fluctuation in the gas-amplification process.

\begin{figure}
 \centering \leavevmode
 \scalebox{0.7}{\includegraphics{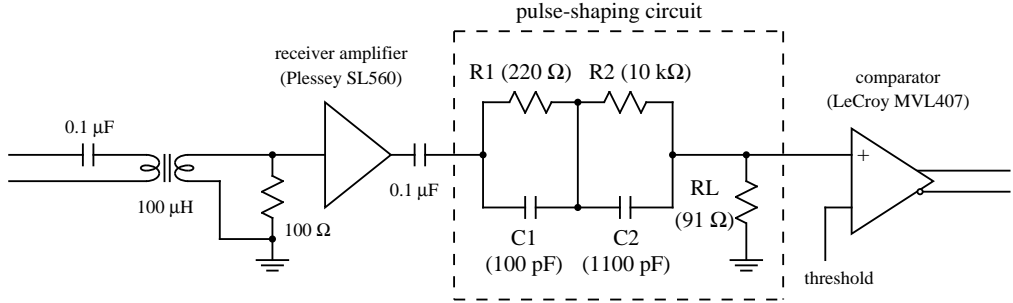}}
 \caption{Schematic diagram of the circuits on the discriminator board, 
to show the added pulse shaping circuit.}
 \label{Fig_shaper}
 \vspace{0.5cm}
\end{figure}

In order to eliminate the tail, 
we added a pulse shaping circuit between the receiver circuit and 
the comparator on the discriminator board. 
A diagram of the added circuit is shown in Fig.~\ref{Fig_shaper}. 
The circuit is a double pole-zero filter, capable of converting 
two poles in the input signal into two new poles.

The parameters of the circuit were determined 
by using signals produced by the 5.9-keV X rays from $^{55}$Fe.
The gas volume of the tube chamber was irradiated through the window 
on the tube wall, 
and the pulse shape at the discriminator input was investigated 
using a digital oscilloscope.

First of all, 
the pulse shape was sampled without adding the shaping circuit, 
and the observed signal tail was approximated by two exponentials.
The circuit parameters were calculated so that the two zeros 
of the circuit should cancel the two time constants (poles) 
of the approximating function.
Another constraint that we applied was that the amplitude 
corresponding to one of the newly produced poles, having a longer 
time constant, should become zero. 

\begin{figure}
 \centering \leavevmode
 \scalebox{0.6}{\includegraphics{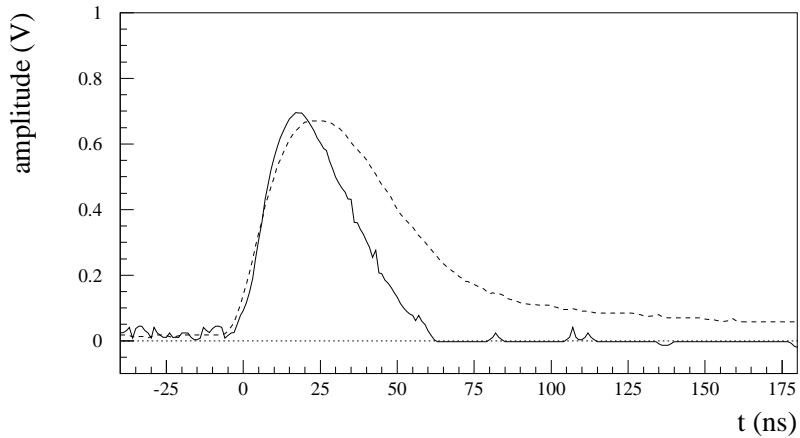}}
 \caption{Measured signal shape for the 5.9-keV X rays from $^{55}$Fe, 
before (dashed) and after (solid) the shaping circuit.
The amplitude of the signal before the circuit is reduced 
so that it becomes comparable with that after the circuit.}
 \label{Fig_pulse}
 \vspace{0.5cm}
\end{figure}

Ideally, a thus-determined circuit should replace a long tail 
in the input signal with one relatively short exponential tail.
However, 
since the approximation with two exponentials has ambiguity, 
fine-tuning was necessary in order to achieve satisfactory 
performance. 
Looking at the resulting pulse shape,
we adjusted the values of two resistors (R1 and R2), 
with capacitors (C1 and C2) and the other resistor (RL) 
fixed to the calculated values.
The optimized parameter values are shown in Fig.~\ref{Fig_shaper}. 
The pulse shapes measured before and after the shaping circuit are 
shown in Fig.~\ref{Fig_pulse}.

\section{Results}

Due to a large fluctuation in the ionization and avalanche processes, 
discriminated signals for cosmic-ray tracks are sometimes separated 
into several fragments.
In the offline analysis, 
successive signals were merged and considered to be one signal 
if the time interval between them (the interval between the trailing 
edge of the preceding signal and the leading edge of the following 
signal) was shorter than 40 ns.

\begin{figure}
 \centering \leavevmode
 \scalebox{0.6}{\includegraphics{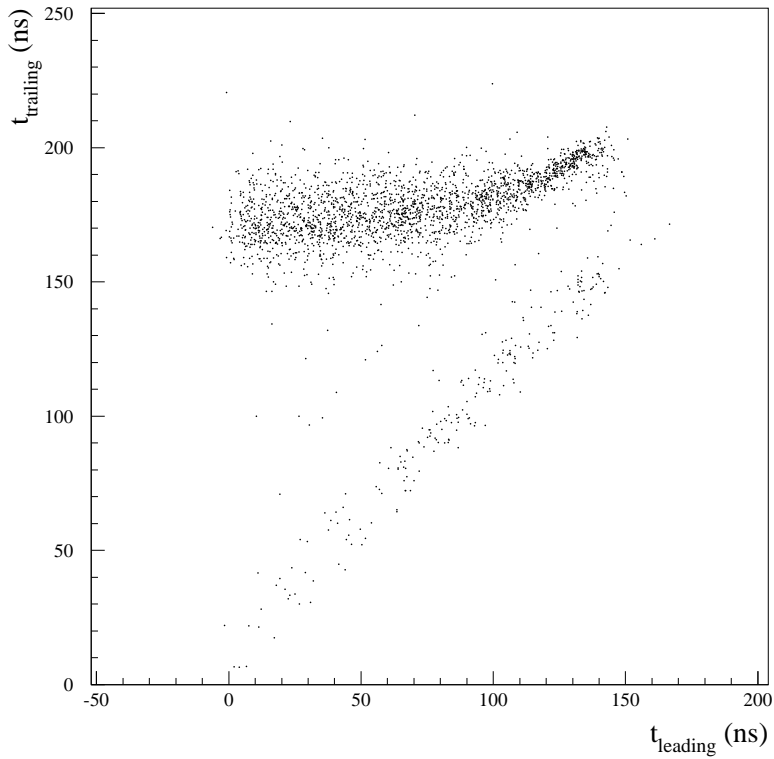}}
 \caption{Scatter plot to show the relation between the leading-edge 
time and the trailing-edge time of the cosmic-ray signals. 
The data were obtained with an anode voltage of 2.0 kV 
and a discriminator threshold of 10 mV.}
 \label{Fig_scatter}
 \vspace{0.5cm}
\end{figure}

Figure \ref{Fig_scatter} shows the relation between the trailing-edge 
time and the leading-edge time of the recorded data.
The data were obtained for an anode voltage of 2.0 kV and 
a discriminator threshold of 10 mV.
The average pulse height for the $^{55}$Fe X rays was 750 mV 
for this anode voltage.
Since about 200 electrons are expected to be produced by this X ray, 
the threshold corresponds to about three-times the average pulse height 
for one ionization electron.

We can see that the trailing edge of the signals exhibits 
a nearly equal time, irrespective of the leading-edge time, 
as expected.
We can also find other interesting and unexpected properties 
in this result: as the leading-edge time becomes larger, 
the trailing-edge time resolution becomes better 
and the average time shifts towards larger values. 
Such properties are expected to emerge as the result of a geometrical 
focusing effect; {\it i.e.}, 
the ionization electron density in the time domain becomes higher 
as the track distance becomes larger.
In a real situation with a finite pulse width, 
this leads to a larger pulse height for signals having larger 
leading-edge times, resulting in a better time resolution 
and a longer delay of the trailing edge.
However, as is shown later, 
this effect is not enough to explain the observed {\it swing-up} 
behavior at large leading-edge times.

\begin{figure}
 \centering \leavevmode
 \scalebox{0.6}{\includegraphics{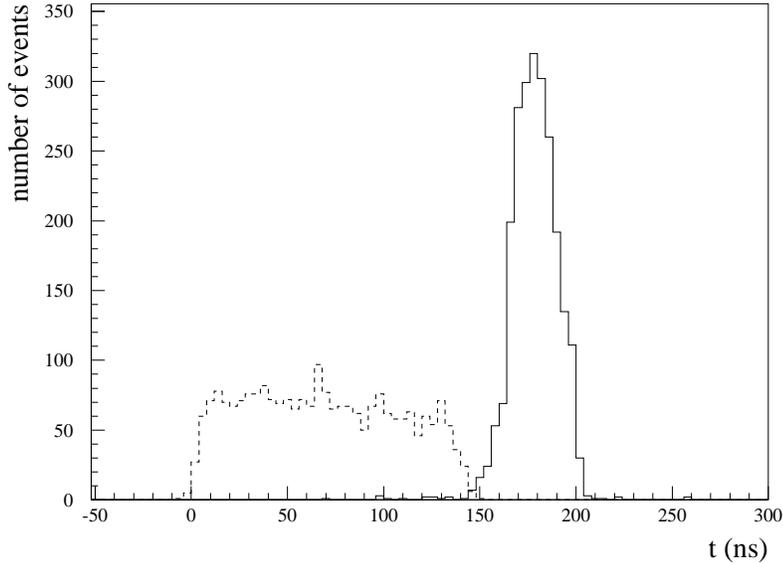}}
 \caption{Projections of the plot in Fig.\ \protect\ref{Fig_scatter}.
The solid histogram shows the distribution of the trailing-edge time, 
and the dashed histogram that of the leading-edge time.
Those data points distributed diagonally 
in Fig.\ \protect\ref{Fig_scatter} are excluded.}
 \label{Fig_time}
 \vspace{0.5cm}
\end{figure}

Along with the dominant data points having a nearly equal 
trailing-edge time, 
we can see some data points distributed diagonally in the plot.
Since the pulse-merging process is applied, 
they are not fragments of wider signals, 
but are isolated narrow signals.
These data points gradually vanish as the threshold voltage is raised.
They are apparently synchronized with the scintillator trigger.
In addition, the frequency of these signals increases 
as the leading-edge time becomes larger, 
This suggests a uniform occurrence of the causal process 
over the chamber gas volume.
These facts indicate that these signals were produced by soft X rays 
coming in association with triggering cosmic-ray muons.
Therefore, they must have nothing to do with the chamber performance 
that we are now interested in.

Projections of the scatter plot are shown in Fig.~\ref{Fig_time}, 
where the data points corresponding to those narrow signals 
described above are excluded.
A flat distribution of the leading-edge time confirms a uniform 
distribution of the cosmic-ray tracks.
The trailing-edge data are concentrated around about 40 ns 
after the maximum leading-edge time 
(the maximum drift time in the ordinary definition).
From a straightforward numerical evaluation, 
we obtained an rms resolution of 12 ns 
for the trailing-edge measurement.

\begin{figure}
 \centering \leavevmode
 \scalebox{0.6}{\includegraphics{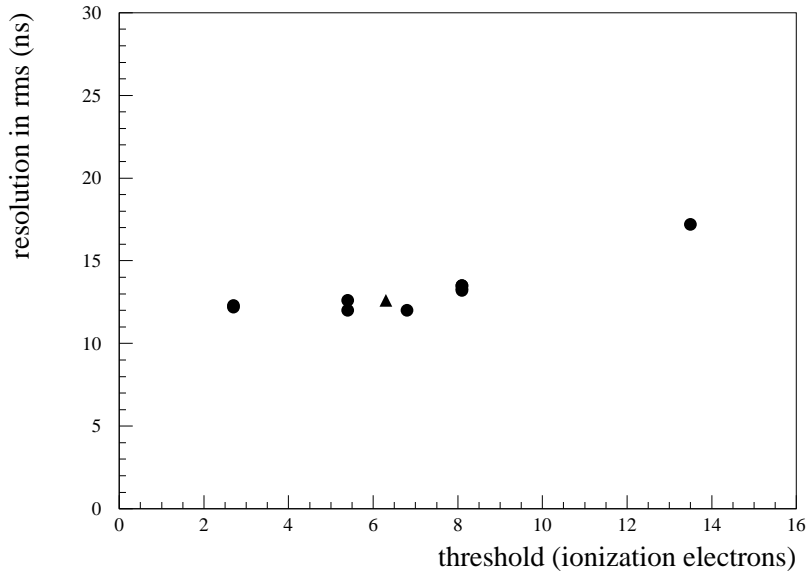}}
 \caption{Trailing-edge time resolution in rms 
as a function of the threshold normalized to the average pulse height 
for one ionization electron. 
The filled circles show the results for an anode high-voltage 
of 2.0 kV. 
The result of an additional measurement at 1.9 kV is plotted 
with a filled triangle.}
 \label{Fig_resolution}
 \vspace{0.5cm}
\end{figure}

The measurement was repeated by varying the threshold voltage.
Figure \ref{Fig_resolution} shows the obtained rms resolution 
as a function of the threshold.
We can observe that the improvement of the resolution 
by lowering the threshold is not significant.
The improvement is limited by the {\it swing-up} behavior seen 
in Fig.~\ref{Fig_scatter}.

\section{Comparison with a simulation}

We developed a simple Monte Carlo simulation,
aiming at understanding the observed properties.
In the simulation, a muon track passing through the chamber gas volume 
leaves ionization electron clusters along its path, 
according to a Poisson distribution with an average frequency of 3.0 
clusters/mm.
The number of electrons composing each cluster is subject to a Poisson 
distribution with an average of 3.0.
The drift time of each electron is determined by the distance 
from the anode wire, assuming a constant drift velocity.
The diffusion during the drift is not taken into account, 
since it is expected to be ineffective, 
compared to the measured time resolution.

\begin{figure}
 \centering \leavevmode
 \scalebox{0.6}{\includegraphics{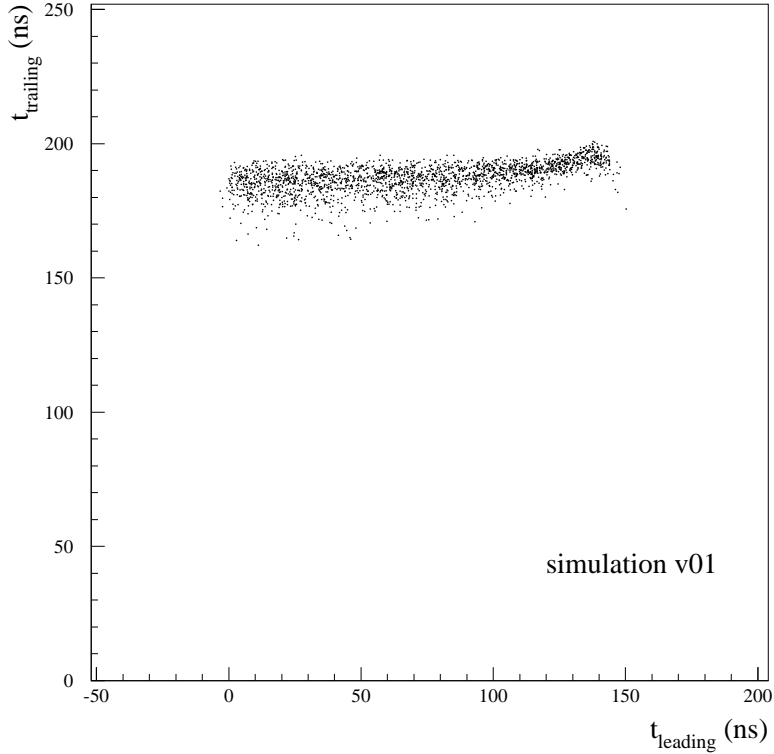}}
 \caption{Relation between the leading-edge time and the trailing-edge 
time obtained from the simplest simulation. 
This is to be compared with the measurement result 
in Fig.\ \protect\ref{Fig_scatter}.}
 \label{Fig_scatterMC}
 \vspace{0.5cm}
\end{figure}

The signal shape is simulated by convoluting the drift time 
distribution with a single-electron response, 
determined from the average pulse shape for $^{55}$Fe X rays.
In the convolution, the pulse height for each electron is varied 
in order to simulate the gas-gain fluctuation.
A Gaussian distribution was assumed for the fluctuation 
in the first version of the simulation.
The leading and trailing-edge times are then determined 
by applying a threshold to the simulated signal shape.

A scatter plot obtained from the simulation, 
which should be compared with Fig.~\ref{Fig_scatter}, 
is shown in Fig.~\ref{Fig_scatterMC},
where the gain fluctuation was assumed to be 30\% 
in the standard deviation.
We can see in this result 
an improvement of the resolution and a shift of the average 
at larger leading-edge times.
However, these variations are apparently less significant 
than those observed in the measurement result.
Namely, the geometrical focusing effect, 
which is automatically included in the simulation, 
is not enough to reproduce the observation.

Aiming at a better reproduction,
we applied several modifications to the simulation. 
The Polya distribution \cite{Polya} was applied to make 
the gas-gain fluctuation more realistic.
The threshold was made to fluctuate according to a Gaussian 
distribution in order to simulate a noise contribution.
The diffusion of drift electrons was taken into account.
Although these modifications could smear the overall time resolution, 
they could never enhance the {\it swing-up} behavior.

As a result of further studies, 
we found that a small overshoot in the shaped signal may produce 
an appreciable {\it swing-up} structure in the leading-trailing 
relation.
The signal shape that we originally used in the simulation does not 
have any overshoot, 
because no significant overshoot was observed in the $^{55}$Fe X-ray 
response.
However, any signal-shape measurement more or less affects 
the circuit property. 
There may be a small, but finite, overshoot in the real situation. 

\begin{figure}
 \centering \leavevmode
 \scalebox{0.6}{\includegraphics{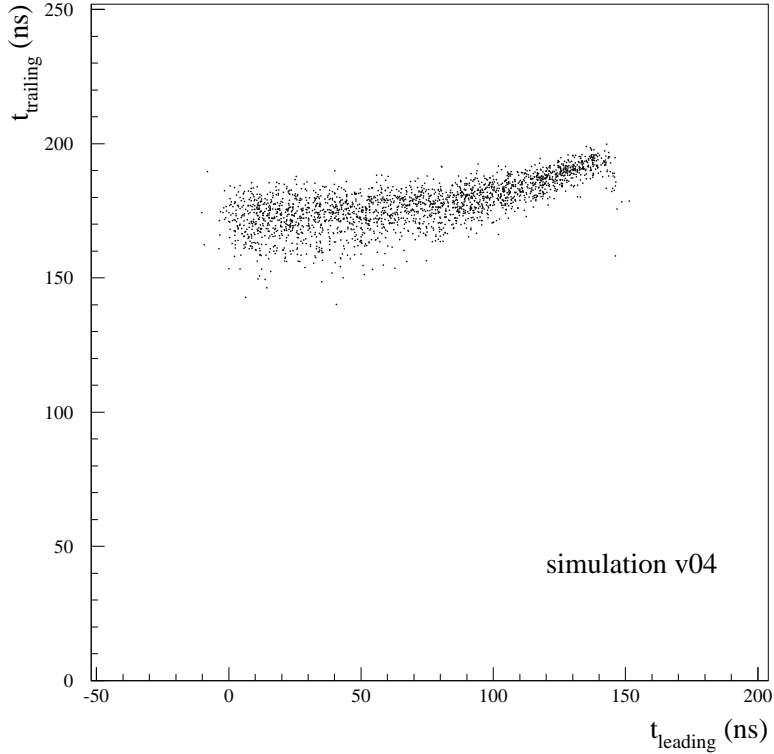}}
 \caption{Leading-trailing relation from a modified simulation, 
where the one-electron response is assumed to have a finite overshoot, 
5\% of the peak pulse height.
Additional modifications concerning the gain and threshold 
fluctuation are also applied.}
 \label{Fig_scatterMC2}
 \vspace{0.5cm}
\end{figure}

Figure~\ref{Fig_scatterMC2} shows the result of a modified simulation 
in which the one-electron response is assumed to have an overshoot 
amounting to 5\% of the peak pulse height.
The overshoot is assumed to produce a baseline shift after the signal;
{\it i.e.}, the overshoot is assumed to have a very long time constant.
The modifications concerning the signal fluctuation mentioned above 
are all applied in this simulation. 
The Polya distribution of a 100\% fluctuation is used for the gain 
fluctuation. 
The standard deviation of the threshold fluctuation is equal to 
the average pulse height for one ionization electron.
The {\it swing-up} behavior 
that we can see in Fig.~\ref{Fig_scatterMC2} looks quite similar 
to that which we observed in the measurement data.

\begin{figure}
 \centering \leavevmode
 \scalebox{0.6}{\includegraphics{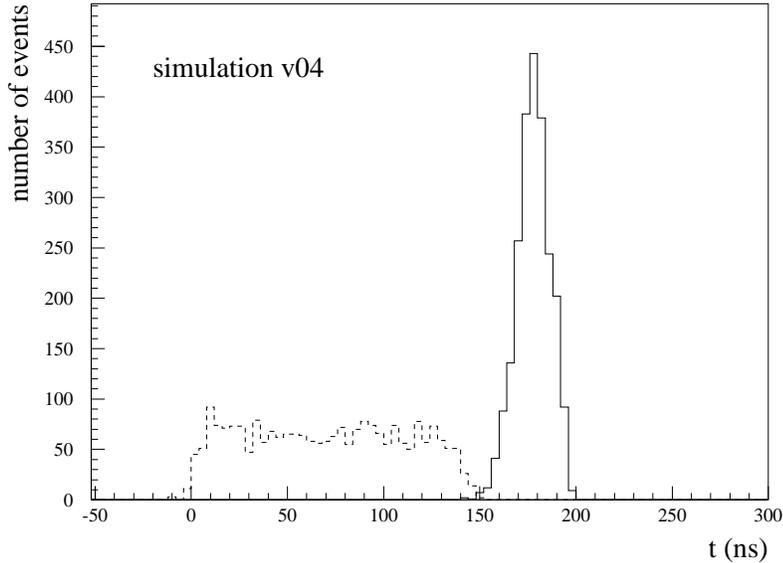}}
 \caption{Projections of the plot 
in Fig.\ \protect\ref{Fig_scatterMC2}, 
to be compared with the measurement result 
in Fig.\ \protect\ref{Fig_time}.}
 \label{Fig_timeMC}
 \vspace{0.5cm}
\end{figure}

Projections of the plot are shown in Fig.~\ref{Fig_timeMC}.
The trailing-edge time distribution has an rms resolution of 8.7 ns.
This is still significantly smaller than that of the measurement result, 
suggesting that a further fine structure of the signal tail 
or certain phenomena not taken into account in the simulation 
({\it e.g.}, the $\delta$-ray emission from the tube wall 
\cite{VENUS-MU}) may be effective. 
Note that non-proportional effects, such as the space-charge effect, 
must not be significant, because a measurement with the anode HV 
lowered to 1.9 kV, for which the gas gain is reduced to about one half, 
gives a comparable result, as shown in Fig.~\ref{Fig_resolution}. 

\section{Conclusions}

A cosmic-ray test was carried out to investigate the feasibility of 
an idea of filtering tube drift-chamber signals 
at high-rate experiments, based on a trailing-edge time measurement.
A small tube chamber filled with a popular chamber gas, Argon/Ethane 
(50/50), at the atmospheric pressure was used for the test.
The leading and trailing-edge times of the signals 
for cosmic-ray tracks were measured simultaneously, 
using a TDC module employing the TMC LSI. 

Applying a simple pole-zero filter circuit for signal shaping, 
we achieved a trailing-edge time resolution of 12 ns in rms 
with a realistic, or rather moderate, setting of 
the discriminator threshold.
Measurements with a resolution of this level will be
very useful at future high-rate experiments.
In the case of LHC, 
such a measurement will allow bunch-crossing identification 
with a tolerance of two or three crossings 
without any significant loss of signals.

In the measurement data, 
we observed an unexpected correlation between the leading and 
trailing edges at large leading-edge times.
This correlation limited the achieved trailing-edge time resolution.

From a simulation study, 
we found that a small overshoot in the signal tail can produce
a correlation quite similar to that which we observed. 
If this is truly the reason, 
the resolution may be further improved with better signal shaping. 
Meanwhile, in the case of long tube chambers,
this effect may seriously limit the achievable resolution, 
since the signal shape varies according to the signal transmission 
length.
Further studies are necessary to confirm these arguments.

\begin{ack}
The authors wish to thank Yasuo Arai and Masahiro Ikeno for their help 
in preparing and maintaining the data-acquisition system. 
Nobuhiro Sato and Takeo Konno are acknowledged for their contribution 
in preparing the test setup.
\end{ack}


\end{document}